\documentclass{ws-p8-50x6-00}

\usepackage{hyperref}
\usepackage[hyperpageref]{backref}

\def\anti{\overline}

\begin{document}

\title{Measuring the relative CP-even and CP-odd Yukawa couplings of a Higgs boson at a muon-collider Higgs factory}

\author{B. Grzadkowski and J. Pliszka}

\address{Institute of Theoretical Physics, Warsaw University, 
ul. Hoza 69, PL-00-681 Warsaw, Poland\\
E-mails: bohdang@fuw.edu.pl,pliszka@fuw.edu.pl}

\author{J. F. Gunion}

\address{Davis Institute for High Energy Physics, 
UC Davis, CA, USA\\
E-mail: jfgucd@physics.ucdavis.edu}  

\maketitle

\abstracts{
We study a possibility of a measurement of muon Yukawa couplings in 
s-channel Higgs boson production at a muon collider with transversely 
polarized beams. We investigate sensitivity to the relative size
of the CP-odd and CP-even muon Yukawa couplings. 
Provided the event rate observed justify the operation of the
$\mu^+\mu^-$ Higgs boson factory, we have found that
polarization degree 40\% is sufficient to resolve the CP nature
of a single resonance as well as disentangle it from two overlapping
CP conserving resonances.
}

It is believed that the scalar sector is an inherent component of the
theory of elementary interactions and, one or more
physical Higgs, or Higgs-like bosons will, sooner or later, be discovered.
Since the CP nature of Higgs bosons is a model dependent feature\cite{hhg} its determination
would not only provide information concerning the mechanism of CP violation
but would also restrict possible extensions of the Standard Model (SM) of electroweak interactions 
and therefore reveal the structure of fundamental interactions beyond the SM.
A muon collider with transversely polarized beams is the
only place where CP properties of a second generation fermion Yukawa coupling
could be probed. This is the subject of our
more complete analysis\cite{hcpmupmum} which is summarized in 
this talk.\footnote{Presented by J.~Pliszka.}
We follow the line of our previous works\cite{gghgp}
where we have tried to unveil the CP-nature of Higgs bosons
in a model-independent way.

The attractive possibility of  s-channel Higgs boson production
at a muon collider has been discussed before\cite{bbgh}
together with the possibility of the measurement of CP violation in the 
muon Yukawa couplings\cite{bbgh,soni,hcpmupmum}.
The latter is based on the fact that 
in any muon collider design\cite{pstareport,euroreport}
there is a natural beam polarization on the order of 20\%\cite{rajat}
that allows for a rare possibility of direct Higgs boson production 
with known polarization of initial state particles.

The cross section for the Higgs boson resonance production:$\mu^+\mu^-\!\rightarrow\!R $ 
depends on transverse $P_T^\pm$ and longitudinal $P_L^\pm$
beam polarizations and $\anti\mu r \exp( i \delta\gamma_5)\mu$ 
muon Yukawa coupling in the following way:
\begin{equation}
\sigma_S(\zeta)=
\sigma_S^0 \left[1+P_L^+P_L^-+P_T^+P_T^-\cos(2\delta+\zeta)\right]
\label{sigform}
\end{equation}
where $\zeta$ is the angle in the transverse plane between 
beam polarizations  and  $\sigma_S^0$ is the unpolarized cross section.
We stress that only the transverse polarization term is sensitive to 
$\delta$ of the muon Yukawa coupling. Since it is proportional
to the product of the transverse polarizations,
it is essential to have large $P_T^+$ and $P_T^-$, as obtained
by applying stronger cuts while selecting
muons from decaying pions (which, however, causes a reduction of luminosity).
To compensate, more 
intensive proton source or ability to repack muon bunches will be needed.
Another  speculative option, would be high, up to 50\%,
polarization obtained by a phase-rotation technique\cite{kaplan}
which would lead to less luminosity reduction.

While varying $\zeta$  one can observe a maximum
at $\zeta=-2\delta$ and a minimum at $\zeta=\pi-2\delta$.
Thus, studying $\zeta$ dependence is essential for resolving the $\delta$
value. A muon collider offers the unique possibility of a setup which
in a natural way provides a scan over different $\zeta$
values. We will not discuss this option here. 
Our results will correspond to a configuration with four
fixed $\zeta$ values: 0,90,180 and 270.
Even though this cannot be accomplished experimentally,
due to the spin precession in the accelerator ring, 
it can be well approximated by a simple but realistic setup\cite{hcpmupmum}
that yields the same results as the fixed $\zeta$ analysis
at the expense of 50\% luminosity increase.

In order to illustrate the ability to reject different Higgs boson
CP scenarios we can assume that the measured data is mimicked by 
the SM Higgs boson. \begin{table}[t]
\caption{1 and 3 $\sigma$ exclusion limits on $\delta$ for $b\bar{b}$ 
final state for various luminosity and polarization configurations.
Beam energy spread 
0.003\% and $b\bar{b}$ tagging efficiency  $54\%$ have  been assumed.}
\begin{center}
\begin{tabular}{|c|c|c|c|c|c|}
\cline{3-6}
\multicolumn{2}{c|}{} & 
\multicolumn{2}{|c|}{$m_R$=110 GeV} & 
\multicolumn{2}{|c|}{$m_R$=130 GeV} \\
\hline
P[\%] & L[pb$^{-1}$] & 1$\sigma$& 3$\sigma$& 1$\sigma$& 3$\sigma$\\
\hline
20 & 150 & 0.94 & -- &  -- & -- \\
39 & 75  & 0.30 & 1.14 & 0.41 & --\\
48 & 75  & 0.20 & 0.64 & 0.27 & 0.93 \\
45 & 150 & 0.15 & 0.50 & 0.21 & 0.69\\
\hline
\end{tabular}
\end{center}
\label{tab_results}
\end{table}
For given luminosity $L$ and polarization $P$ we can place
1 and 3 $\sigma$ limits on the $\delta$ value for the observed resonance,
assuming the $\delta=0$ SM is input.
The limits for Higgs boson masses  of 110 and 130 GeV are presented
in table~\ref{tab_results}. For the expected yearly luminosity of
$L=100~{\rm pb}^{-1}$, even several years of running
at the natural 20\% polarization would be insufficient for useful limits. 
However, 1$\sigma$ limits for the $P=39\%$ option (with reduced $L$) 
do give a rough indication of the CP nature of the resonance.
3$\sigma$ limits 
in 110-130 GeV mass range require either $>40\%$ polarization 
or $<50\%$ luminosity loss.
(The requirements are less stringent for a 110 GeV Higgs boson.)
We stress that there is no other way the measurement 
of the muon Yukawa $\delta$ can be done
and that operation in the transverse polarization mode should not
interfere with most of the other studies.
\begin{table}[t]
\caption{Event number pattern for different Higgs models as
a function of $\zeta$, assuming $P_L^\pm=0$ and $P_T^\pm=P$; see
Eq.~(\ref{sigform}).}
\begin{center}
\begin{tabular}{|c|c|c|c|c|}
\hline
\raisebox{0pt}[12pt][7pt]{$(\delta)$} & $\zeta=0$ & 
$\zeta=\pi/2$  & $\zeta=\pi$ & $\zeta=3\pi/2$ \\
\hline
\raisebox{0pt}[10pt][5pt]{$(0)$} & $1+P^2$ & 1 & $1-P^2$ & 1 \\
\raisebox{0pt}[5pt][5pt]{$(\pi/4)$} & 1 & $1-P^2$ & 1 & $1+P^2$ \\
\raisebox{0pt}[5pt][5pt]{$(\pi/2)$} & $1-P^2$ & 1 & $1+P^2$ & 1 \\
\raisebox{0pt}[5pt][5pt]{$(0)+(\pi/2)$ }& 1 & 1 & 1 & 1 \\
\hline
\end{tabular}
\end{center}
\label{patterns}
\end{table}
For a heavier resonance, operation of a muon collider as an s-channel
Higgs boson factory is justified only if the branching ratio
$BR(R\rightarrow \mu^+\mu^-)$ is enhanced. Then, the analysis 
sketched above applies as well.
Such enhancement arises in the Minimal Supersymmetric 
Standard Model (MSSM) at large $\tan\beta$.
If the pseudoscalar mass is large ($m_A >300\textrm{GeV}$), the
$H$ and $A$ masses will be similar. For $\tan\beta>8$ degeneracy
may be such that we will not be able to see two separate peaks
but only a single peak with both CP-even and CP-odd components.
It would be crucial to distinguish such a case from a single
CP-violating Higgs boson which may appear e.g. in MSSM\cite{cp-phases}. 
Table \ref{patterns} illustrates
the very distinct event number pattern as a function of $\zeta$
that would yield the needed discrimination.
The event rate for any single, CP conserving or CP violating Higgs boson,
has a minimum and maximum as a function of $\zeta$. In contrast,
overlapping CP-even and CP-odd resonances
result in a pattern independent of $\zeta$.
We have found that for simple MSSM test cases with $m_A= 300-400$ GeV
and $\tan\beta> 8$ (for which we cannot see separate resonance peaks) 
even natural 20\% polarization
will allow us to distinguish two overlapping resonances from any single one
at more than the 3$\sigma$ level.
Higher polarization will allow for a precise measurement
of the relative contribution from the CP-even and the CP-odd component.

To summarize, we have presented results of 
a realistic study of measuring the
CP properties of the muon Yukawa couplings in 
Higgs boson production at a muon collider with transversely polarized beams.
We have found that transverse polarization is essential for determining
the CP nature of the 
muon Yukawa couplings. In particular, a collider with 
$P\!\approx\!40\%$ and at least 50\% of the 
original luminosity retained 
will ensure that the CP nature of the produced scalar resonance 
will be revealed.
\section*{Acknowledgments}
We thank S. Geer, R. Raja and R. Rossmanith for helpful conversations
on experimental issues.
This work was supported in part by the U.S. Department of Energy,
the U.C. Davis Institute for High Energy Physics, the State Committee for
Scientific Research (Poland) grant No. 2~P03B~014~14 and by Maria
Sklodowska-Curie Joint Fund II 
(Poland-USA) grant No. MEN/NSF-96-252.

\end{document}